\documentclass[prd,aps,11pt,%showpacs,
nofootinbib,tightenlines,superscriptaddress,preprintnumbers]{revtex4}

\def\lsim{\raise0.3ex\hbox{$<$\kern-0.75em\raise-1.1ex\hbox{$\sim$}}}
\def\gsim{\raise0.3ex\hbox{$>$\kern-0.75em\raise-1.1ex\hbox{$\sim$}}}

\usepackage{graphicx}

\newcommand{\beq}{\begin{equation}}
\newcommand{\eeq}{\end{equation}}
\newcommand{\bqa}{\begin{eqnarray}}
\newcommand{\eqa}{\end{eqnarray}}

\begin{document}

\title{Causal Viscous Hydrodynamics for Central Heavy-Ion Collisions}

\preprint{BI-TP 2006/39}\preprint{INT PUB 06-18}

\author{Rudolf Baier}
\affiliation{Fakult\"at f\"ur Physik, Universit\"at Bielefeld, 
D-33501 Bielefeld, Germany}
\author{Paul Romatschke}
\affiliation{Institute for Nuclear Theory, University of Washington,
Box 351550, Seattle WA, 98195, USA}
\date{\today}

\begin{abstract} 
We study causal viscous hydrodynamics in the context of 
central relativistic heavy-ion collisions and provide details
of a straightforward numerical algorithm to solve the hydrodynamic
equations. It is shown that
correlation functions of fluctuations provide stringent test
cases for any such numerical algorithm. Passing these tests,
we study the effects of viscosity on the temperature profile
in central heavy-ion collisions. 
Also, we find that it is possible to counter-act the effects
of viscosity to some extent by re-adjusting the initial conditions.
However, viscous corrections are strongest for high-mass particles,
signaling the breakdown of hydrodynamic descriptions for 
large $\eta/s$.
%
%large mass particles are most affected by viscous corrections,
%indicating a breakdown 
%
%conditions for small $\eta/s$. For $\eta/s=0.3$ or larger, viscous
%corrections to these slopes become large and hydrodynamic descriptions
%seem to break down.
\end{abstract}

\maketitle

\section{Introduction}

Successful fits of ideal hydrodynamics 
to experimental data on several observables~\cite{Teaney:2000cw,Huovinen:2001cy,Kolb:2001qz,Hirano:2002ds,Kolb:2002ve}
 at the highest energies
of the ongoing heavy-ion program 
~\cite{Adcox:2004mh,Back:2004je,Arsene:2004fa,Adams:2005dq} at 
the Relativistic Heavy-Ion Collider (RHIC)
seem to indicate a very small value of the ratio shear
viscosity over entropy.
Since calculations of this ratio in QCD at weak coupling $\alpha_s\ll1$ 
give \cite{Trans1,Trans2}
a numerical value that turns out to be larger by about one order
of magnitude than the conjectured strong coupling value for relativistic quantum
field theories at finite temperature \cite{Kovtun:2004de} (see also
\cite{Janik:2006ft}), 
this has given rise to the idea of a ``strongly-coupled'' quark-gluon
plasma phase~\cite{Shuryak:2004cy,Lee:2005gw,Gyulassy:2004zy,Heinz:2005zg,Muller:2006ee,Blau:2005pk,Riordan:2006df}.

However, up to now calculations in the regime of strong coupling
are limited to theories which possess a gravity dual theory, 
such as ${\mathcal N}=4$ Super Yang-Mills Theory \cite{Klebanov:2000me}.
So far, no such dual theory has been discovered for QCD, 
meaning that the main available tool to study dynamical processes
in QCD are based on weak-coupling approaches (although 
lattice-based techniques for making quantitative measurements of
near-equilibrium quantities may be available soon 
\cite{Karsch:1986cq,Nakamura:2004sy}).

Given that the numerical value of the QCD coupling $\alpha_s$
within the range of temperatures applicable for RHIC
is assumed to be close to the range $\alpha_s=0.2-0.4$, 
one observes that while this value is not very small, it is not
very large either. Thus, although it would be of great interest to
have results for QCD at very strong coupling, there is at least
some hope that existing weak-coupling techniques might actually offer
a description of RHIC physics that is not inferior to still-to-be-discovered 
QCD strong-coupling techniques (or likewise, extrapolating existing
strong-coupling results from theories (very) different than QCD).

Along these lines, it has recently been discovered
that non-Abelian plasma instabilities
\cite{Mrowczynski:2005ki} create turbulent color magnetic fields
\cite{Arnold:2005ef, Arnold:2005qs} 
that may induce a very small effective (``anomalous'')
shear viscosity coefficient \cite{Asakawa:2006tc,Asakawa:2006jn},
without invoking strong coupling effects.
Within the initial conditions obtained in the color-glass-condensate
model \cite{Iancu:2003xm} it is, however, unclear whether this effect 
is relevant for present RHIC energies \cite{Romatschke:2005pm,
Romatschke:2006wg}.

Regardless of these issues, it is important to note that so far
the ratio of shear viscosity over entropy density for RHIC energies
is fairly unconstrained. While the general trend of viscous 
corrections to ideal hydrodynamics has been studied by Teaney
\cite{Teaney}, a dynamical implementation of viscous hydrodynamics
and comparison to experimental data is still lacking.
This is partly due to the fact that the ``simplest'' form
of viscous hydrodynamics, the relativistic Navier-Stokes equations,
are be-riddled by acausality problems and instabilities \cite{Hisc}.
Therefore, there has been recent interest in so-called second-order
(Israel-Stewart\cite{IS}) theories \cite{Prakash:1993bt,Muronga,
Muronga:2004sf,Heinz:2005bw,Baier:2006um,Baier:2006sr,Tsumura:2006hn,Koide:2006ef} which are, however, of more complicated structure
than the Navier-Stokes equations.

Specifically, it seems that adapting existing numerical hydrodynamic
solvers to treat Israel-Stewart theory for all but the simplest geometries
is a non-trivial task. It might therefore be worth-wile to 
devise completely new algorithms that are more suitable (or at least
simpler) than present hydrodynamic solvers. Along these lines,
in this work we present a straightforward algorithm for solving Israel-Stewart
viscous hydrodynamic equations for geometries that are longitudinally 
expanding, are space-time rapidity independent and have radial symmetry
and thus should be well suited to describe viscous hydrodynamics
of central collisions at RHIC and in the future at the 
Large Hadron Collider (LHC).

Our work is organized as follows: in section \ref{sec:two} we review
the equations of causal viscous hydrodynamics and present evidence
that our numerical algorithm reproduces the ideal hydrodynamic behavior
in the limit of a small ratio of viscosity over entropy, as it should.

In section \ref{sec:three}, we present a more involved test
that is based on measuring correlation functions of small fluctuations
and comparing to analytic results.

In section \ref{sec:four}, our results 
for the temperature evolution and particle spectra in 
relativistic causal viscous hydrodynamics are presented and we 
give our conclusions in section \ref{sec:five}.

\section{Setup and Comparison with Ideal Hydrodynamics}
\label{sec:two}

The basic equations of causal viscous hydrodynamics
that we choose to study are given by \cite{Baier:2006um}
\bqa
(\epsilon+p)D u^\mu&=&\nabla^\mu p-\Delta^\mu_{\nu} \nabla_\sigma
\Pi^{\nu \sigma}+\Pi^{\mu \nu} D u_\nu\, ,
\label{2.1}\\
D \epsilon &=& - (\epsilon+p) \nabla_\mu u^\mu+\frac{1}{2}\Pi^{\mu \nu}
\langle\nabla_\nu u_\mu\rangle\, ,
\label{2.2}\\
\tau_{\Pi} \Delta^\mu_\alpha \Delta^\nu_\beta D \Pi^{\alpha \beta}
+\Pi^{\mu \nu}&=&\eta  \langle\nabla^\mu u^\nu\rangle
- 2 \tau_{\Pi} \Pi^{\alpha (\mu}\omega^{\nu)}_{\ \alpha}\, ,
\label{baseq}
\eqa
where $\epsilon,p$ are the energy density and pressure, respectively,
$u^\mu$ is the flow four-velocity that obeys $u_\mu u^\mu=1$ and
$\Pi^{\mu \nu}$ is the shear tensor that fulfills
$u_\mu \Pi^{\mu \nu}=0=\Pi^\mu_\mu$ and characterizes 
the viscous deviations in the energy momentum tensor,
\beq
T^{\mu \nu}=(\epsilon+p)u^\mu u^\nu-p g^{\mu \nu}+\Pi^{\mu \nu}.
\label{Tdef}
\eeq 
Furthermore, $\eta$ and $\tau_\Pi$ are the 
shear viscosity coefficient and relaxation time that 
are related by $\frac{\eta}{\tau_\Pi}=\frac{2 p}{3}$ 
in weakly-coupled QCD \cite{Baier:2006um} and
the remaining definitions are
\bqa
&d_\mu u^\nu \equiv \partial_\mu u^\nu+\Gamma_{\alpha \mu}^{\nu}
u^\alpha,\qquad
D\equiv u_\mu d^\mu,\qquad
\nabla^\mu \equiv \Delta^{\mu \nu} d_\nu,&
\nonumber\\
&\Delta^{\mu\nu}\equiv g^{\mu \nu}-u^\mu u^\nu, \qquad
\omega^{\mu \nu}=\Delta^{\mu \alpha} \Delta^{\nu \beta}
\frac{1}{2}\left(d_\beta u_\alpha-d_\alpha u_\beta\right),&\nonumber\\
&\langle A_\mu B_\nu\rangle \equiv  A_\mu B_\nu+A_\nu B_\mu
-\frac{2}{3} \Delta_{\mu \nu} A_\alpha B^\alpha,\qquad
(A_\mu,B_\nu) \equiv 
\frac{1}{2}\left(A_\mu B_\nu+A_\nu B_\mu\right),&
\label{symbdef}
\eqa
where $\Gamma_{\alpha \mu}^\nu$ are the Christoffel symbols.
As outlined in the introduction, we will be interested in
systems which are rapidity-invariant and have radial symmetry,
therefore we choose to work in 
co-moving and radial coordinates $\tau,r,\phi,\eta$ with
the relations $\tau=\sqrt{t^2-z^2}$, $r^2=x^2+y^2$, $\tan{\phi}=y/x$
and $\eta={\rm atanh} (z/t)$.
The only non-vanishing fluid velocity
components are then $u^\tau$ and $u^r$ with the relation
$u^\tau=\sqrt{1+(u^r)^2}$ and neglecting gradients in $\phi$ and
$\eta$ we find for the above equations
\bqa
(\epsilon+p)D u^\tau&=&\left(1-(u^\tau)^2\right)\left(\partial_\tau p
-d_\nu \Pi^\nu_\tau\right)
-u^\tau u^r \left(\partial_r p-d_\nu \Pi^\nu_r\right)\nonumber\\
(\epsilon+p)D u^r&=&-u^\tau u^r\left(\partial_\tau p-d_\nu \Pi^\nu_\tau\right)
-\left(1+(u^r)^2\right) \left(\partial_r p-d_\nu
\Pi^\nu_r\right)\nonumber\\
D \epsilon &=&-(\epsilon+p) \theta +\frac{1}{2}\left(
-\Pi^r_r (1-v^2)^2\langle\nabla^r u^r\rangle
-r^2\, \Pi^\phi_\phi \langle\nabla^\phi u^\phi\rangle
-\tau^2\,\Pi^\eta_\eta \langle\nabla^\eta u^\eta \rangle
\right)\nonumber\\
- d_\nu \Pi^\nu_\tau &=& v^2 \partial_\tau \Pi^r_r
+v \partial_r \Pi^r_r+\Pi^r_r\left(
\partial_\tau v^2+\partial_r v+\frac{v^2}{\tau}
+\frac{v}{r}\right)+\frac{1}{\tau}\Pi^\eta_\eta
\nonumber\\
d_\nu \Pi^\nu_r &=& v\, \partial_\tau \Pi^r_r+
\partial_r \Pi^r_r+
\Pi^r_r \left(\partial_\tau v+\frac{v}{\tau}+\frac{2-v^2}{r}\right)
+\frac{1}{r} \Pi^\eta_\eta
\nonumber\\
\tau_{\Pi} D \Pi^\eta_\eta +\Pi^\eta_\eta&=& 
-\eta\, \tau^2\, \langle \nabla^\eta
u^\eta \rangle\nonumber\\
\tau_{\Pi} D \Pi^r_r
+\Pi^r_r&=&-\eta\,\langle\nabla^r u^r\rangle 
+2 u^r \tau_\Pi \left(\Pi^r_\tau D u^\tau + \Pi^r_r D
u^r\right)
\nonumber
\eqa
\bqa
&\langle \nabla^r u^r \rangle =-2 \partial_r u^r-2 u^r D u^r +
\frac{2}{3} \left(1+(u^r)^2\right) \theta,\qquad
r^2 \langle \nabla^\phi u^\phi \rangle =
-2 \frac{u^r}{r}+\frac{2}{3}\theta,&
\nonumber\\
&\tau^2 \langle \nabla^\eta u^\eta \rangle =
-2 \frac{u^\tau}{\tau}+\frac{2}{3}\theta,\qquad
\theta = \partial_\tau u^\tau+\partial_r u^r+\frac{u^\tau}{\tau}
+\frac{u^r}{r}&\,
\label{alleqs}
\eqa
where $v=u^r/u^\tau$, $\Pi^r_\tau=-v\, \Pi^r_r$, 
$\Pi^\phi_\phi=-\Pi^\eta_\eta-(1-v^2) \Pi^r_r$ and 
here $D=u^\tau \partial_\tau+u^r \partial_r$.
This system of equations has to be closed by providing an equation of
state, e.g. $\epsilon=\epsilon(p)$.

Clearly, it is possible to use the relation $u^\tau=\sqrt{1+(u^r)^2}$
to eliminate either $u^\tau$ or $u^r$ from the above
equations. Defining $\gamma=u^\tau=(1-v^2)^{-1/2}$ 
one obtains
\bqa
\left[\gamma^4(\epsilon+p)-(1-v^2 \gamma^2) \Pi^r_r\right]
\partial_\tau v&=&
-\gamma^2 (\partial_r + v \partial_\tau) p
+(\partial_r+v\partial_\tau) \Pi^r_r
-\left[\gamma^4 v (\epsilon+p)+\gamma^2 v \Pi^r_r\right] \partial_r v
\nonumber\\
&&+\left(\frac{v}{\tau}+\frac{2}{r}\right)\Pi^r_r-\gamma^2
\left(\frac{v}{\tau}- \frac{1}{r}\right) \Pi^\eta_\eta\nonumber\\
\partial_\tau \epsilon&=&
-\left[ (\epsilon+p) \gamma^2-\Pi^r_r\right] v \partial_\tau v
-v \partial_r \epsilon-(\epsilon+p) \left[\gamma^2 \partial_r
v+\frac{1}{\tau}+\frac{v}{r}\right]\nonumber\\
&&+\Pi^r_r\left[\partial_r v-\frac{v}{\gamma^2 r}\right]
-\Pi^\eta_\eta \left[\frac{v}{r}-\frac{1}{\tau}\right].
\label{veqs}
\eqa

However, in the code 
we prefer to keep both $u^\tau$ and $u^r$, solving equations for them
independently so that a non-trivial consistency check on the numerics
is provided by monitoring the deviation of $(u^\tau)^2-(u^r)^2$ from unity.

\subsection{Discretization}

We use discretized space-time and compute differentials
of a function $f(\tau,r)$ as finite differences,
\beq
\partial_r f(\tau,r) = \frac{f(\tau,r+a)-f(\tau,r-a)}{2a},\qquad
\partial_\tau f(\tau,r) = \frac{f(\tau+\delta
\tau,r)-f(\tau,r)}{\delta \tau},
\eeq
where $a,\delta \tau$ are the spatial and temporal
lattice spacings, respectively. The boundaries are taken care of by
using one-sided derivatives.

Provided a starting condition at time $\tau=\tau_0$ for the variables
$u^\tau,u^r,\epsilon,\Pi^r_r,\Pi^\eta_\eta$ one can then integrate the
set of equations (\ref{alleqs}) forward in time.
The virtue of this approach is that one immediately obtains the results
for the fluid velocities etc. rather than having to perform the ``usual''
hydrodynamic algorithm (transforming to the calculational frame,
integrating equations, transforming back).
The drawback is that in Eqs.~(\ref{alleqs}), time derivatives of the 
above variables are still coupled (e.g. the first equation of
Eqs.~(\ref{alleqs}) contains both $\partial_\tau u^\tau$ and
$\partial_\tau p$). However, since all time derivatives enter only
linearly this can be rectified by making use of a linear equation
solver so that e.g. $\partial_\tau u^\tau=f(\tau,r)$ which can be
directly integrated using the above discretization.

In practice, this works as follows: from the first equation in
Eqs.(\ref{alleqs}) we pick out the coefficients of the time
derivatives $\partial_\tau u^\tau,\partial_\tau u^r,\partial_\tau p$
and label them as $a_{00},a_{01},a_{02}$, respectively. The remaining part of
the equation, which contains no time derivative, is called $b_0$.
Thus, it becomes of the form
\beq
a_{00} \partial_\tau u^\tau + a_{01} \partial_\tau u^r + a_{02} \partial_\tau p
= b_0.
\eeq
Note that in order to obtain this form we have expanded out the 
derivatives $d_\nu \Pi^\nu_\tau,d_\nu \Pi^\nu_r$ by using the relevant
equations in (\ref{alleqs}).
A similar procedure for the second and third equation of
(\ref{alleqs}) leaves us with
\bqa
a_{10} \partial_\tau u^\tau + a_{11} \partial_\tau u^r + a_{12} \partial_\tau p
&=& b_1\nonumber\\
a_{20} \partial_\tau u^\tau + a_{21} \partial_\tau u^r + a_{22} \partial_\tau p
&=& b_2.
\eqa
With $\delta_j=\partial_\tau(u^\tau,u^r,p)$, 
these three equations may be written in matrix form as $a_{ij} \delta_j =
b_i$, which has a solution unless ${\rm det}\, a_{ij}=0$.
Numerically, this matrix equation 
is readily solved using a standard linear-equations solver, so $\delta_j$ is
known explicitly and may be used to finally compute $\partial_\tau
\Pi^\eta_\eta$ and $\partial_\tau \Pi^r_r$. This completes the setup of
our algorithm\footnote{A version of the code, written in reasonably
well documented C, may be obtained upon request from paulrom@physik.uni-bielefeld.de}.

\subsection{Testing the Code -- Ideal Hydrodynamics}

\begin{figure}
\vspace{0.6cm}
\begin{center}
\includegraphics[width=0.4\linewidth]{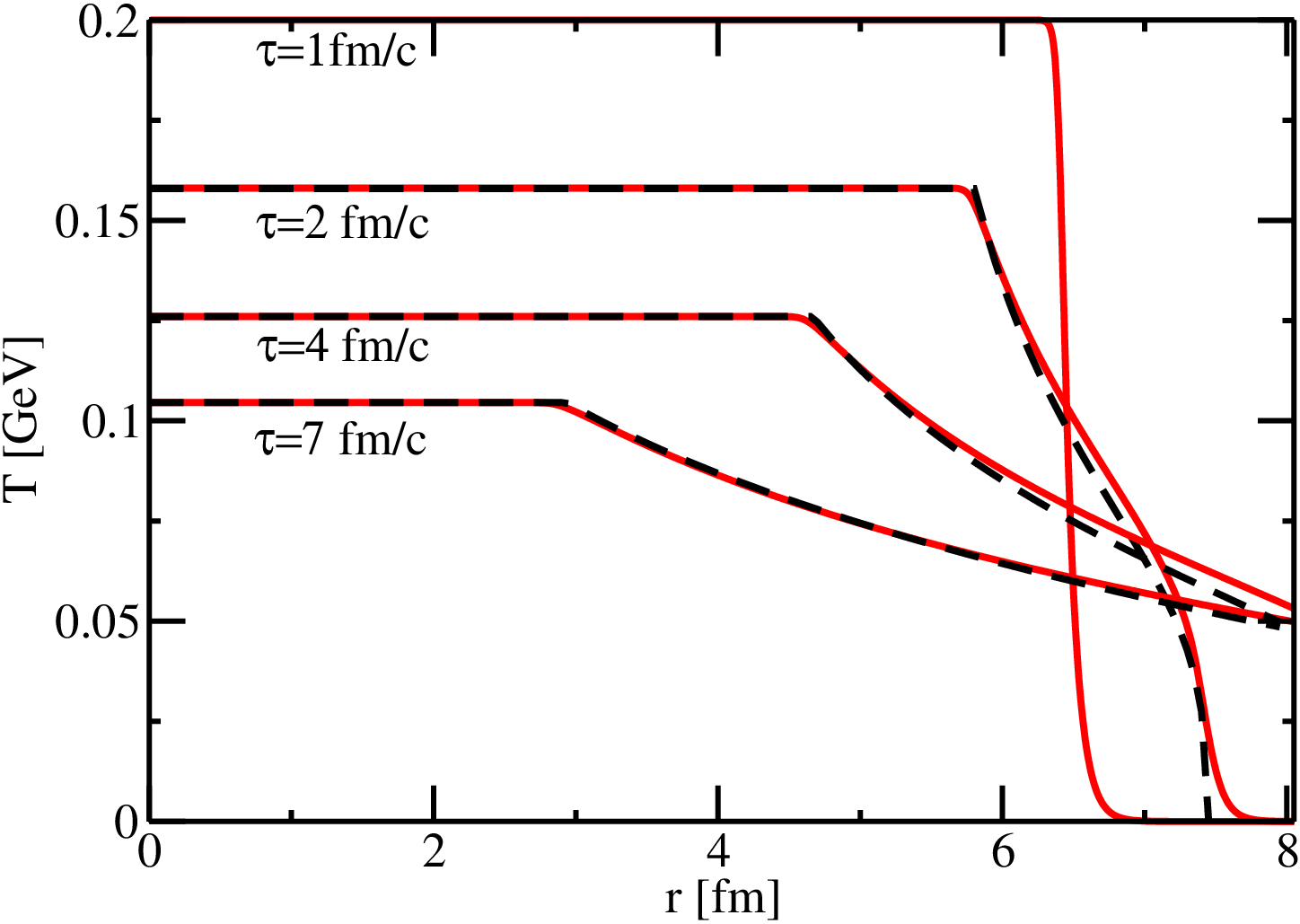}
\hspace*{1cm}
\includegraphics[width=0.4\linewidth]{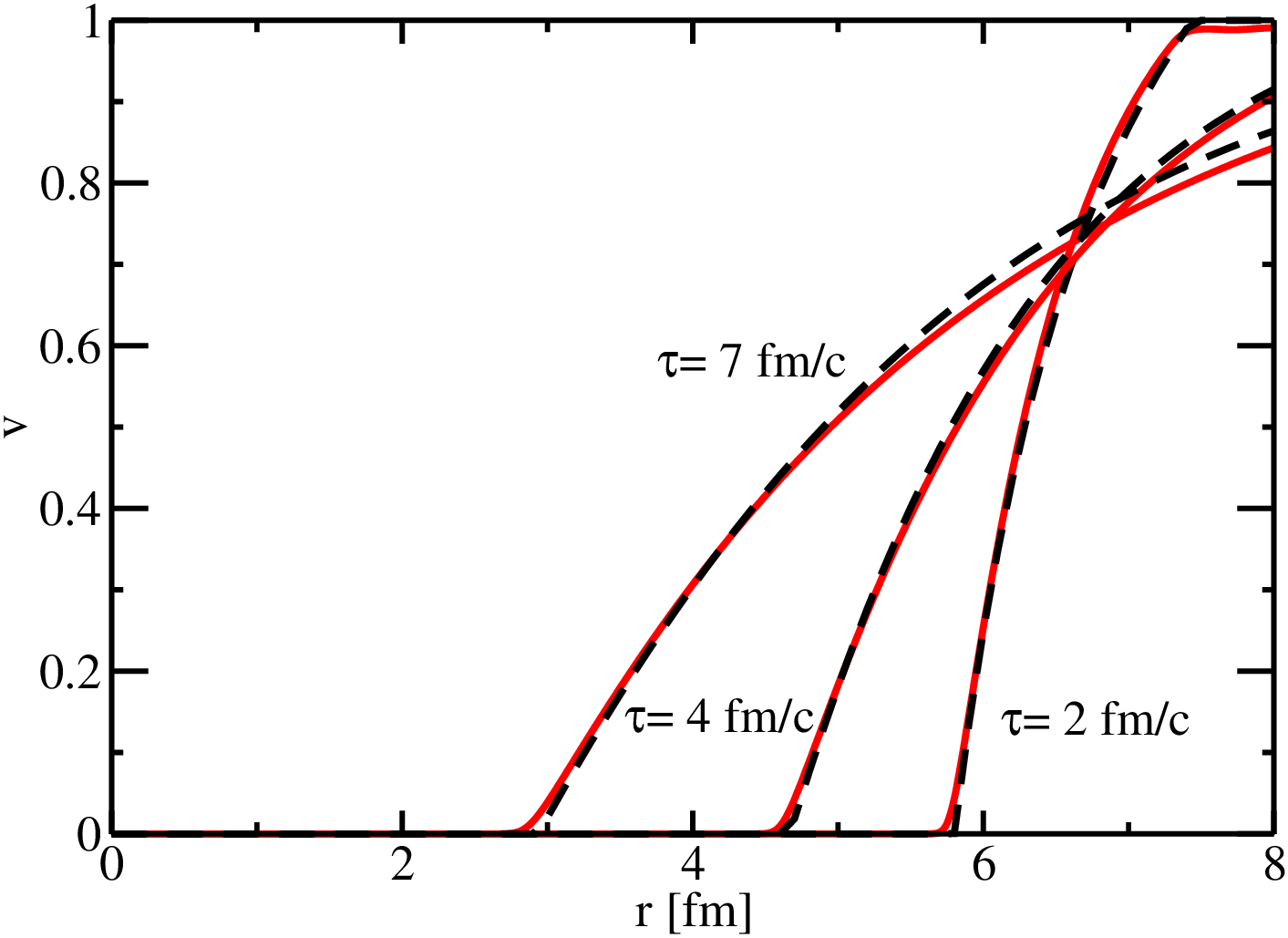}
\end{center}
%\vspace{.6cm}
\caption{Comparison between numerical results and the analytic
approximation Eq.(\ref{Baymres}) (full and dashed lines,
respectively) for the evolution of the temperature (left figure) and
the velocity $v=\frac{u^r}{u^\tau}$ (right figure). The slight disagreement
between numerical and analytical results is due to the approximations
involved in the analytic solution and has the same sign and size
as in the original work \cite{Baym:1984sr}.}
\label{fig:htest1}
\end{figure}

As a first test, we run our numerical code for a very small value
of viscosity, $\eta/s=10^{-4}$ and compare our results to ideal
hydrodynamics. Our problem of choice is to start with a configuration
for the energy density
\beq
\epsilon(r,\tau_0)=\frac{\epsilon_0}{1+\exp{[(r-R)/\sigma]}},
\label{WSp}
\eeq
where $R=6.4$ fm can be thought of as the ``radius'' of a nucleus,
and $\epsilon_0$ is such that the temperature (assuming an ideal
gluon gas) is $T_0=0.2$ GeV at $r=0$.
The parameter $\sigma$ is in principle arbitrary, but we choose it to be
\hbox{$\sigma=0.02$ fm} in order to have a very steep fall of the energy density
near $r \simeq R$. Choosing the ideal equation of state $\epsilon(p)=3
p$ for which the speed of sound squared $c_s^2=\frac{1}{3}$, we can
then compare the time evolution of temperature and velocity to the analytic
solution
\bqa
T_{\rm Baym}(r,\tau)&=&T_0 e^{-c_s \alpha_R(r,\tau-\tau_0)}\left(\frac{\tau_0}{\tau}\right)^{
c_s^2\left[1+(1-c_s v_R(r,\tau-\tau_0))^{-1}\right]/2}\nonumber\\
v_{\rm Baym}(r,\tau)&=&{\rm tanh}\left[\alpha_R(r,\tau-\tau_0)+\frac{c_s^2}{2}
\left(\frac{v_R(r,\tau-\tau_0)}{1-c_s v_R(r,\tau-\tau_0)}\right)
\ln\left(\frac{\tau-\tau_0}{\tau_0}\right)\right]\nonumber\\
\alpha_R(r,t)&=&\left\{\begin{array}{r@{\quad:\quad}l}
0& r < R -c_s t\\
-\frac{1}{2}\ln\left(\frac{t-r+R}{t+r-R}
\, \frac{1-c_s}{1+c_s}\right)
 & R -c_s t < r < R+t\\
\infty & r > R+t
\end{array}\right.\nonumber\\
v_R(r,t)&=&\left\{\begin{array}{r@{\quad:\quad}l}
0 & r < R -c_s t\\
\frac{r-R+c_s t}{t+c_s(r-R)}& R -c_s t < r < R+t\\
1 & r > R+t\end{array}\right.
\label{Baymres}
\eqa
from Baym et al. \cite{Baym:1984sr}. Results are shown in 
Fig.\ref{fig:htest1}, where it can be seen that -- within the errors
of the approximate analytic solution -- the numerical solution agrees with
the results Eq.(\ref{Baymres}).

\section{Fluctuations and Linearized Viscous Hydrodynamics}
\label{sec:three}

Motivated by cosmology where one can actually observe correlations
of density fluctuations in the early universe
\cite{Peebles,WMAP,Mukhanov}, we study radial fluctuations of the energy
density $\epsilon$, the flow velocity $v$ and the 
shear tensor $\Pi^{\mu \nu}$ around a background
solution $\epsilon_0,u^r_0,\Pi^{\mu \nu}_0$ such that
\beq
\epsilon(\tau,r)=\epsilon_0(\tau)+\delta \epsilon(\tau,r), \qquad
v(\tau,r)=\delta v(\tau,r), \qquad \Pi^{\mu \nu}=\Pi^{\mu\nu}_0
(\tau)+\delta \Pi^{\mu \nu}(\tau,r) 
\eeq
where the background solution obeys the equations \cite{Muronga,Lallouet:2002th}
\beq
\partial_\tau \epsilon_0=-\frac{\epsilon_0+p_0}{\tau}+\frac{1}{\tau}
\Pi^\eta_{\eta,\, 0} \qquad
\partial_\tau\Pi^\eta_{\eta,\, 0}=-\frac{1}{\tau_\Pi} \Pi^\eta_{\eta,\, 0}
+\frac{8 p_0}{9 \tau}
\label{Backgroundeq}
\eeq
and we remind that $\frac{\eta}{\tau_\Pi}=\frac{2 p_0}{3}$.

In what follows, we will assume that the
fluctuations around the background are small so we keep 
only terms linear in the 
perturbations (``linearized hydrodynamics'', 
c.f. \cite{kadanoff,Casalderrey-Solana:2004qm}). Note, however, that 
we keep the full non-trivial time dependence of the background,
which to the best of our knowledge has not been done before in
the context of heavy-ion collisions even in the case of ideal hydrodynamics.

To slightly simplify the discussion, we want to assume in this section
that $\tau_\Pi$ is constant with respect to time (which consequently requires
a time-dependent ratio of $\eta/s$), whereas in other sections
of this work $\tau_\Pi$ will be time-dependent. 
To linear order in the perturbations one is thus left with the set
of coupled partial differential equations
\bqa
\left[(\epsilon_0+p_0-\Pi^r_{r,0})\partial_\tau+c_s^2 \partial_\tau \epsilon_0
-\left(\partial_\tau+\frac{1}{\tau}\right) \Pi^r_{r,0}+\frac{1}{\tau}\Pi^\eta_{\eta,0}\right]
\delta v+c_s^2\partial_r \delta \epsilon-\left[\partial_r+\frac{2}{r}\right]
\delta \Pi^r_r-\frac{1}{r}\delta \Pi^\eta_\eta &=&0\nonumber\\
\left[(\epsilon_0+p_0)\left(\partial_r+\frac{1}{r}\right)-\Pi^r_{r,0}
\left(\partial_r-\frac{1}{r}\right)+\frac{1}{r}\Pi^\eta_{\eta,0}\right]
\delta v+\left[\partial_\tau+\frac{1+c_s^2}{\tau}\right] \delta \epsilon
-\frac{1}{\tau} \delta \Pi^\eta_\eta&=&0\nonumber\\
\frac{4}{9} p_0 \left[\partial_r+\frac{1}{r}\right]
\delta v-\frac{8}{9} \frac{c_s^2}{\tau} \delta \epsilon+\left[\partial_\tau+
\frac{1}{\tau_\Pi}\right] \delta \Pi^\eta_\eta&=&0\nonumber\\
\frac{4}{9} p_0 \left[-2\partial_r+\frac{1}{r}\right]\delta v
+\frac{4}{9} \frac{c_s^2}{\tau} \delta \epsilon
+\left[\partial_\tau+\frac{1}{\tau_\Pi} \right]\delta \Pi^r_r&=&0.\quad\quad
\eqa
These can be further simplified by noting that 
for the initial condition $\Pi^{\mu \nu}_0=0$ and no radial flow
one has $\Pi^r_{r,0}=\Pi^\phi_{\phi,0}$ and as a consequence of $\Pi^\mu_{\mu,0}=0$
the relation
$\Pi^r_{r,0}=-\frac{1}{2}\Pi^\eta_{\eta,0}$ holds for all $\tau$.

Usually one would do a space-like Fourier transform to get rid of the
space-like derivatives. Due to our choice of coordinates, however, this
is obviously not possible. However, upon introducing
\beq
\delta \bar{\Pi}=\left(\partial_r+\frac{2}{r}\right)\delta \Pi^r_r+\frac{1}{r}
\delta \Pi^\eta_\eta
\eeq
we can achieve the same goal by doing a so-called
Bessel-Fourier transform,
\bqa
&\delta v(\tau,r) = \int_0^{\infty} d\kappa J_1(\kappa r) 
\delta \tilde v(\tau,\kappa), \qquad
\delta \epsilon(\tau,r) = \int_0^{\infty} d\kappa J_0(\kappa r) 
\delta \tilde \epsilon(\tau,\kappa),&\nonumber\\
%\label{BFT}
%\eeq
%\beq
&\delta \bar \Pi(\tau,r) = \int_0^{\infty} d\kappa J_1(\kappa r) 
\delta \tilde \Pi(\tau,\kappa), \qquad
\delta \Pi^\eta_\eta(\tau,r) = \int_0^{\infty} d\kappa J_0(\kappa r) 
\delta \tilde \Pi^\eta_\eta(\tau,\kappa)&,
\label{BFT}
\eqa
where the property of the Bessel functions $J_n$
\beq
\int_0^\infty dr\, r J_n(\kappa r) J_n(\kappa^\prime
r)=\frac{\delta(\kappa-\kappa^\prime)}{\kappa} 
\label{BesselU}
\eeq
can be used to invert the above transformations.
Using an ideal equation of state, $p_0=c_s^2 \epsilon_0$
and the Eqs.~(\ref{Backgroundeq}) to remove explicit
time derivatives on $\epsilon_0$ and $\Pi^\eta_{\eta,0}$ we
thus find
\bqa
\left[(\epsilon_0+p_0+\frac{1}{2}\Pi^\eta_{\eta,0})\partial_\tau
+c_s^2 \partial_\tau \epsilon_0
+\frac{1}{2}\left(\partial_\tau+\frac{3}{\tau}\right) \Pi^\eta_{\eta,0}\right]
\delta \tilde v-\kappa c_s^2\delta \tilde \epsilon-\delta \tilde \Pi
&=&0\nonumber\\
\left[\epsilon_0+p_0+\frac{1}{2}\Pi^\eta_{\eta,0} 
\right] \kappa
\delta \tilde v+\left[\partial_\tau+\frac{1+c_s^2}{\tau}\right] \delta 
\tilde\epsilon
-\frac{1}{\tau} \delta \tilde \Pi^\eta_\eta&=&0\nonumber\\
\frac{4}{9} p_0 \kappa
\delta \tilde v-\frac{8}{9} \frac{c_s^2}{\tau} \delta \tilde \epsilon
+\left[\partial_\tau+\frac{1}{\tau_\Pi}\right] \delta \tilde \Pi^\eta_\eta&=&0\nonumber\\
\frac{8}{9} p_0 \kappa^2
\delta \tilde v
-\frac{4}{9} \kappa \frac{c_s^2}{\tau}\delta \tilde \epsilon
+\left[\partial_\tau+\frac{1}{\tau_\Pi} \right]\delta \tilde \Pi&=&0.
\label{ansol2a}
\eqa

\subsection{Sonic peaks in Ideal Hydrodynamics}

Upon first taking the limit $\tau_\Pi\rightarrow 0$ and then setting
$\Pi^{\mu \nu}_0$ as well as $\delta \Pi^{\mu \nu}$ to zero we recover
the equations for fluctuations in ideal hydrodynamics, which together
with $\epsilon_0\propto \tau^{-1-c_s^2}$ require
\beq
\left[\partial_\tau^2-\frac{c_s^2}{\tau} \partial_\tau+\frac{c_s^2}{\tau^2}+c_s^2
\kappa^2\right]  \delta \tilde v(\tau,\kappa)=0
\eeq
(and a similar differential equation for $\delta \tilde \epsilon$).
As can be quickly verified, the solutions to the linearized ideal hydrodynamic
equations then become
\bqa
\delta v(\tau,r)&=&\int_0^\infty d\kappa J_1(\kappa r)
\tau^{(1+c_s^2)/2} \left[A(\kappa)\, J_{(-1+c_s^2)/2}(\kappa c_s \tau)
+B(\kappa)\, Y_{(-1+c_s^2)/2}(\kappa c_s \tau)\right],
\label{vsol}\nonumber\\
\frac{\delta \epsilon(\tau,r)}{\epsilon_0(\tau)}&=&-\frac{1+c_s^2}{c_s}
\int_0^\infty d\kappa J_0(\kappa r)
\tau^{(1+c_s^2)/2} \left[A(\kappa)\, 
J_{(1+c_s^2)/2}(\kappa c_s \tau)
+B(\kappa)\, Y_{(1+c_s^2)/2}(\kappa c_s \tau)\right],
\label{epssol}
\eqa
where $J$ and $Y$ are both Bessel functions of the first kind and 
$A,B$ are constants of integration.

As initial conditions at the
starting time $\tau=\tau_0$
we choose for simplicity $\delta v(\tau_0,r)=0$
and random noise for $\delta \epsilon(\tau_0,r)$ with a correlation
function
\footnote{
Let $f^{(i)}(r)$ be the $i^{\rm th}$ configuration of an observable
$f$. The correlation function $<f(r) f(r^\prime)>$ is then defined 
as 
$<f(r) f(r^\prime)> \equiv \lim_{N\rightarrow \infty} \frac{1}{N} 
\sum_{i=1}^{N} f^{(i)}(r) f^{(i)}(r^\prime)$.}
\beq
\epsilon_0^{-2} <\delta \epsilon(\tau_0,r) \, \delta \epsilon(\tau_0,r^\prime)>
= \Delta^2 \frac{\delta(r-r^\prime)}{r},
\eeq
which is the polar-coordinate equivalent of a 
Gaussian distribution with standard deviation $\Delta$.

This initial condition has the advantage that it implies
\beq
<A(\kappa) A(\kappa^\prime)> = \kappa \delta(\kappa-\kappa^\prime)
\frac{\Delta^2 c_s^2}{\tau_0^{1+c_s^2} (1+c_s^2)^2}
\left[J_{(1+c_s^2)/2}(\kappa c_s \tau_0)-Y_{(1+c_s^2)/2}(\kappa c_s \tau_0)
\frac{J_{(-1+c_s^2)/2}(\kappa c_s \tau_0)}{Y_{(-1+c_s^2)/2}(\kappa c_s \tau_0)}\right]^{-2},
\eeq
such that
\beq
\epsilon_0^{-2}(\tau)<\delta \epsilon(\tau,r)\,\delta \epsilon(\tau,r^\prime)>
=\int_0^\infty d\kappa\, \kappa\, J_0(\kappa r) J_0(\kappa r^\prime)
f(\kappa,\tau,\tau_0)
\eeq
where
\beq
f(\kappa,\tau,\tau_0)=\left(\frac{\tau}{\tau_0}\right)^{1+c_s^2} \Delta^2
\frac{\left[J_{(1+c_s^2)/2}(\kappa c_s \tau) Y_{(-1+c_s^2)/2}(\kappa c_s
\tau_0)
-Y_{(1+c_s^2)/2}(\kappa c_s \tau)
J_{(-1+c_s^2)/2}(\kappa c_s \tau_0)\right]^{2}}
{\left[J_{(1+c_s^2)/2}(\kappa c_s \tau_0)Y_{(-1+c_s^2)/2}(\kappa c_s
\tau_0)
-Y_{(1+c_s^2)/2}(\kappa c_s \tau_0)
J_{(-1+c_s^2)/2}(\kappa c_s \tau_0)\right]^{2}}.
\label{ffuncres}
\eeq

Despite its ugly appearance, this is a nice result since 
for large $\kappa$ we find
\beq
f(\kappa,\tau,\tau_0)\rightarrow
\left(\frac{\tau}{\tau_0}\right)^{c_s^2} \Delta^2 
\cos^2{(\kappa c_s (\tau-\tau_0))},
\eeq
which are just the sonic peaks that one can also derive in cosmology.

This result can serve as a stringent test on the numerical 
algorithm used to solve the hydrodynamic equations as the
position of the maxima and minima of $f$ as a function of $\kappa$ 
are very sensitive to the speed of sound. In what follows, we 
thus initialize our numerical algorithm with precisely the same
initial conditions as discussed above and then
measure the correlation function
$\epsilon_0^{-2}(\tau)<\delta \epsilon(\tau,r)\,\delta
\epsilon(\tau,r^\prime)>$ to extract $f(\kappa,\tau,\tau_0)$
by using Eq.(\ref{BesselU}) to integrate out\footnote{
Since we solve the hydrodynamic equations on a lattice, in practice we
do the integrations by summing over all lattice sites. Furthermore,
momenta $\kappa=\frac{\pi k}{N a}$ are also discrete, where $a$ is the
spatial lattice spacing, N the number of lattice sites and 
$k$ is a positive integer smaller than $N/2$.} both $r,r^\prime$,
\beq
\int_0^\infty r\, dr \int_0^\infty r^\prime\, dr^\prime
J_0(\kappa r) J_0(\kappa^\prime r^\prime)
\epsilon_0^{-2}(\tau)<\delta \epsilon(\tau,r)\,\delta \epsilon(\tau,r^\prime)>
=\frac{\delta(\kappa-\kappa^\prime)}{\kappa}
f(\kappa,\tau,\tau_0).
\label{inversion}
\eeq
To maximize the signal, we pick $\kappa=\kappa^\prime$ which
is regular on the lattice we use to solve the hydrodynamic equations. 

\begin{figure}
\vspace{0.6cm}
\begin{center}
\includegraphics[width=0.4\linewidth]{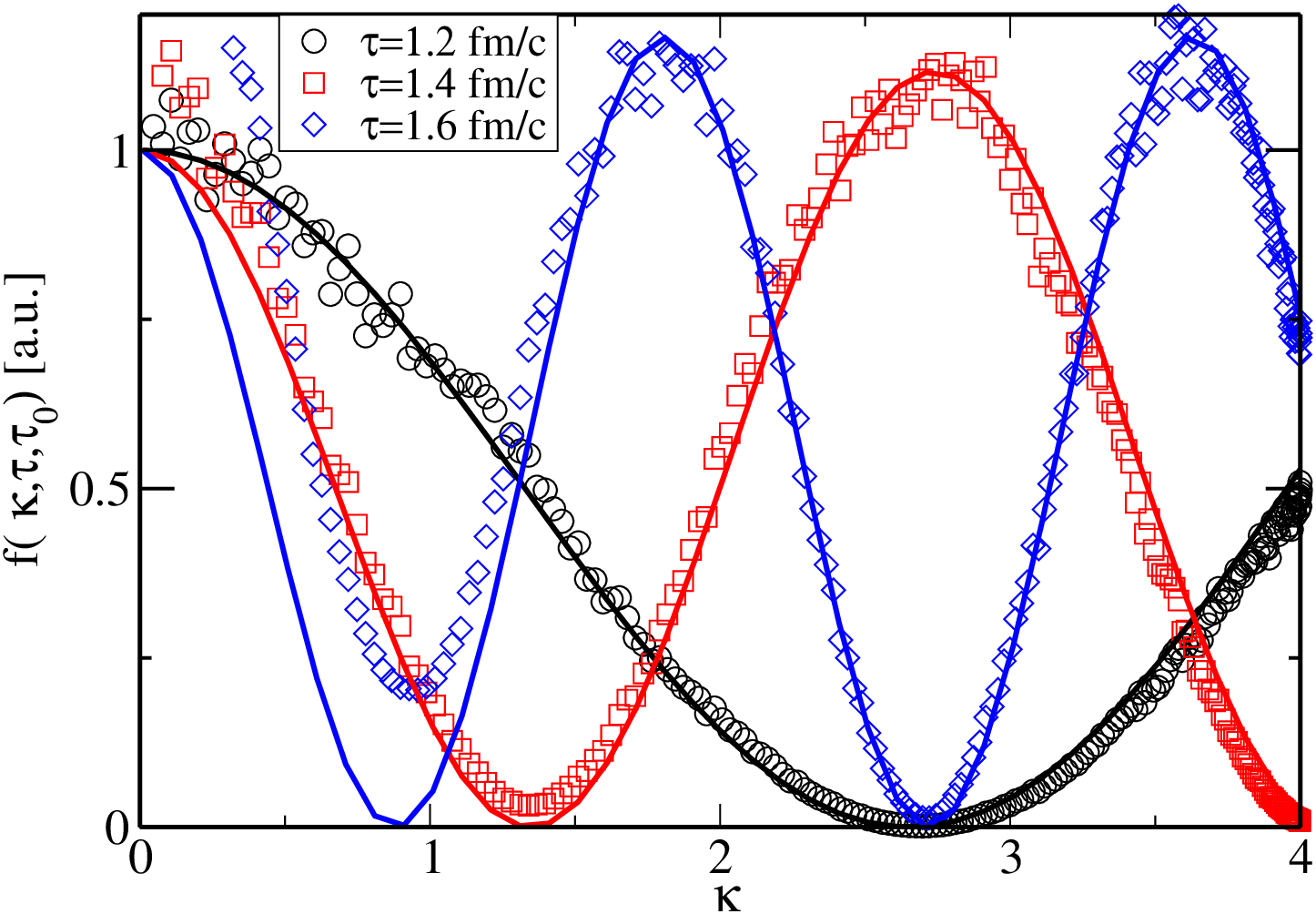}
\hspace*{1cm}
\includegraphics[width=0.4\linewidth]{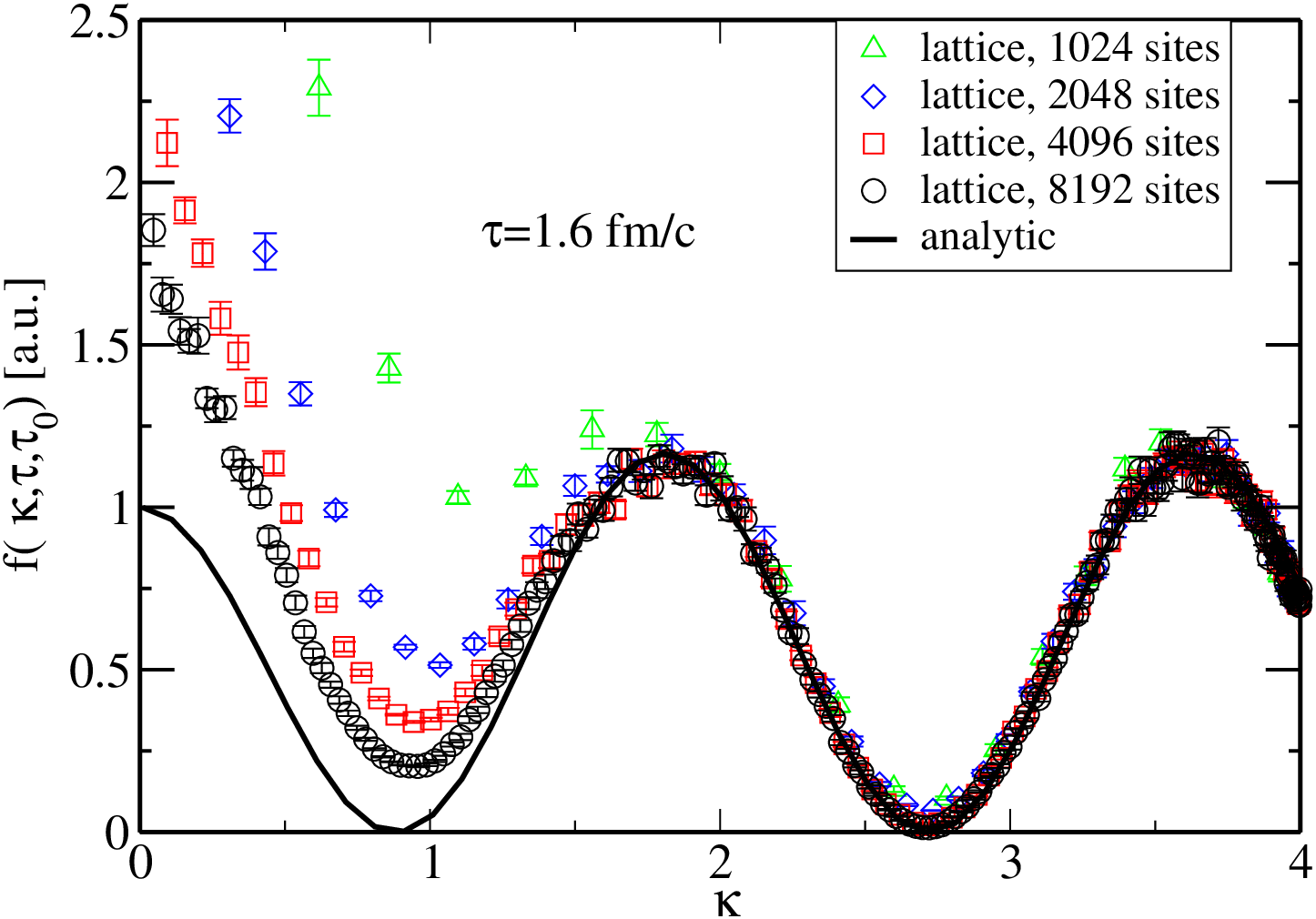}
\end{center}
%\vspace{.6cm}
\caption{Left figure: The correlation function 
$f(\kappa,\tau,\tau_0)$ from solving
hydrodynamic equations on a lattice 
(symbols, see text for details)
for $\tau_0=1 $ fm/c and different $\tau$ compared to the analytic 
result (full lines). As one can see, the agreement between
the analytic result and the measured correlation function is 
very good in general (a slight difference e.g. in the speed of sound
would be clearly visible by a shift in the minima/maxima of $f$).
Note that at later times, a discrepancy at low momenta $\kappa$ develops.
This is probably a lattice artifact since
increasing the simulated volume reduces the 
discrepancy (right figure).}
\label{fig:fcs}
\end{figure}

In Fig.\ref{fig:fcs} we show the result
for\footnote{
Note that the lattice dispersion relation $\kappa=a^{-1} \sin{(\pi k/N)}$
has been used to convert to continuum values.}
$f(\kappa,\tau,\tau_0)$ 
obtained on a lattice with lattice spacing $a=0.25$ GeV$^{-1}$ and $N=8192$
sites and $\eta/s=10^{-4}$, 
ensemble-averaged over 100 configurations and coarse grained
in $\kappa$, for three different times $\tau$.
Up to three sonic peaks can be nicely distinguished
and the comparison with the analytic result Eq.(\ref{ffuncres})
indicates that our code 
indeed accurately solves the ideal hydrodynamic equations with 
the ``correct'' speed of sound.
There is, however, a slight discrepancy between the numerical measured
correlation function and its analytic result at small $\kappa$ 
and later times: presumably, this is due to the fact that on the lattice,
only a finite number of momenta can be simulated and thus 
the inversion formula Eq.(\ref{inversion}) holds only approximately.
Indeed, the second part of Fig.\ref{fig:fcs} shows that this discrepancy
can be systematically reduced by going to larger lattice volumes.
Since the solution of the viscous hydrodynamic equations themselves
do not depend on relations such as Eq.(\ref{inversion}), this discrepancy
should not be mistaken as a failure of the algorithm to correctly treat
low momentum modes.

\subsection{Sonic Peaks in Viscous Hydrodynamics}

Treating the set of equations (\ref{ansol2a}) in their
full generality we were unable to find analytic solutions
like those obtained in the previous subsection. However,
since together with Eq.(\ref{Backgroundeq}) these are just
a set of six coupled ordinary differential equations they
readily lend themselves to numerical solutions, which we nevertheless
want to refer to as ``analytic'' since they are completely independent
of our numerical algorithm to solve the hydrodynamic equations.

\begin{figure}
\vspace{0.6cm}
\begin{center}
\includegraphics[width=0.4\linewidth]{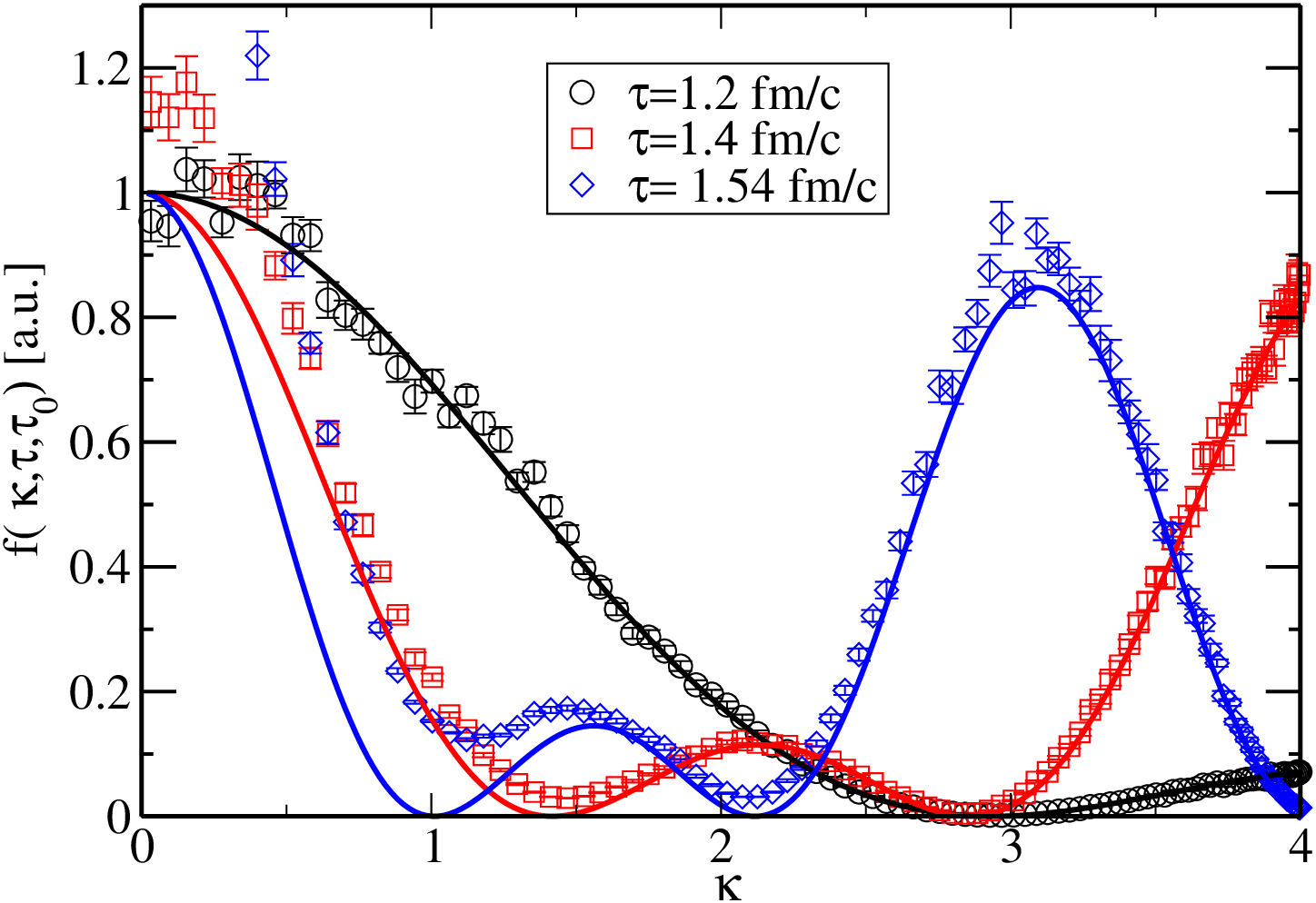}
\hspace*{1cm}
\includegraphics[width=0.4\linewidth]{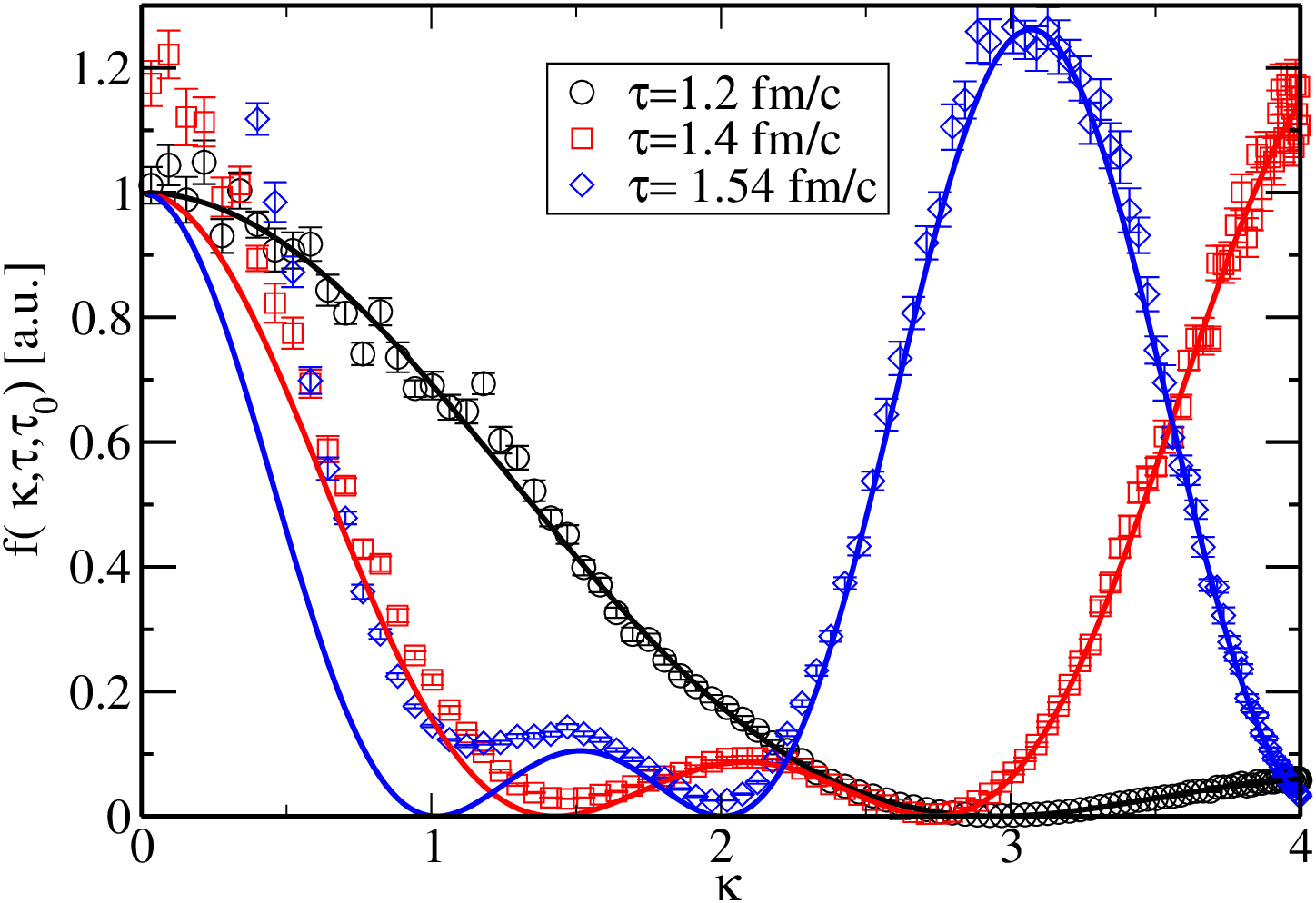}
\end{center}
\caption{Comparison for $f(\kappa,\tau,\tau_0)$ from 
solving hydrodynamic equations on a lattice (symbols, $\eta/s=0.1$ (left)
and $\eta/s=0.3$ (right), respectively) with
4096 sites and lattice spacing $a=0.25\, {\rm GeV}^{-1}$ and 
``analytic'' solution (full lines) of Eqs.(\ref{ansol2a}).}
\label{fig:fcs2}
\end{figure}

With the same initial conditions
as in the previous subsection one therefore obtains an ``analytic''
solution of the correlation function $f(\kappa,\tau,\tau_0)$, 
with both expansion and viscosity included.
In Fig.\ref{fig:fcs2}, this ``analytic'' solution is again compared 
to the correlation function obtained by solving the hydrodynamic
equations on a lattice (for $\eta/s=0.1$ and $\eta/s=0.3$, respectively).
Similarly to the case of vanishing viscosity, 
we find that there is very good general agreement between the 
measured (ensemble averaged and coarse-grained) correlation function 
and the ``analytic'' result, except for later times and small momenta
$\kappa$, where lattice artifacts seem to be accumulating.

Since also in this case the position and width of the sonic 
peaks are very sensitive to the value of $\eta/s$ and the 
speed of sound, we argue that the good agreement between
measured and ``analytic'' correlation functions is a strong
indication that our numerical code is indeed correctly solving
the second order viscous hydrodynamic equations.

Finally, we want to point out that fluctuation measurements 
may also help to constrain the value of $\eta/s$ from RHIC data,
as has been recently suggested \cite{Gavin:2006xd}.

\section{Causal Viscous Hydrodynamics with Transverse Flow}
\label{sec:four}

Let us now study the effects of viscosity on quantities of interest
for heavy-ion collisions. For simplicity, we will assume that
the radial energy density profile is given by Eq.(\ref{WSp}), where
we take $R_0=6.4$ fm and $\sigma=0.54$ fm which has been used before
for ideal hydrodynamic calculations \cite{Baym:1984sr}.
The constant $e_0$ is chosen such that we have an initial temperature
$T_0$ at $r=0$.
In accordance with ideal hydrodynamic studies in their simplest form
we assume that at the
time when we start applying our hydrodynamic description ($\tau=\tau_0$),
the system does not have any transverse flow already, so $v=0$.
Furthermore, since this is still an exploratory study, 
we pick a simplistic equation of state, $p=c_s^2 \epsilon$
with constant speed of sound $c_s^2=1/3$.

This set of initial conditions would be sufficient to determine the 
subsequent dynamics fully in the case of ideal hydrodynamics. Including
the effects of viscosity requires that we pick a specific value of the 
ratio $\eta/s$ and also provide initial conditions for the two independent
components of $\Pi^{\mu \nu}$ at $\tau=\tau_0$.
Maybe the simplest choice would be to  assume -- like in ideal
hydrodynamics -- that the system for some reason happens to be in 
equilibrium at $\tau=\tau_0$, such that ``accidentally'' $\Pi^{\mu \nu}=0$.
This choice probably highlights best the difference of viscous hydrodynamics
to ideal hydrodynamics, since one starts from the same initial condition, 
so we will use it as our initial condition in the following.

However, there are other ``sensible'' choices of initial conditions 
that might be more relevant for real heavy-ion 
collisions in the future. E.g. within the Color-Glass-Condensate model
in its simplest form (the McLerran-Venugopalan model \cite{MV}), the
system does not have any longitudinal dynamics, so after times 
$\tau>Q_s^{-1}$ where $Q_s$ is the saturation scale, the system essentially
has zero longitudinal pressure \cite{Krasnitz:1999wc,Lappi:2003bi}.
In the local rest frame $u^r=u^\phi=u^\eta=0$, so with $c_s^2=1/3$, 
Eq.(\ref{Tdef}) would imply
\beq
\Pi^\eta_\eta=p, \quad \Pi^r_r=- \frac{p}{2}
\label{CGC}
\eeq

Finally, going beyond the McLerran-Venugopalan model to include 
so called next-to-leading order corrections of gluon production
\cite{Gelis:2006yv,Gelis:2006cr}
one has to take into account the effect of rapidity fluctuations
and full three-dimensional gauge field dynamics. This has recently
been shown to trigger instabilities \cite{Romatschke:2005pm}, 
leading to the generation
of non-zero longitudinal pressure \cite{Romatschke:2006nk}
at $\tau> Q_s^{-1}$. Pending
%an analytical 
%an understanding of 
the result using the correct rapidity fluctuation spectrum
\cite{francoistalk},
the initial condition for $\Pi^{\mu \nu}$ is expected to
lie somewhere in-between the two cases discussed above.

\subsection{Temperature Profile in Viscous Hydrodynamics}

Choosing the initial condition $\Pi^{\mu\nu}=0$ at
$\tau=\tau_0$, we can investigate the changes of the temperature profile
from the ideal hydrodynamic behavior due to dynamical viscous effects.
In Fig.{\ref{fig:tprof} we show the temperature as a function of
the radius for $T_0=0.36$ GeV at $\tau_0=1$ fm/c,
but different values of $\eta/s$, calculated on a lattice with
$512$ sites and $a=0.25$ $\rm GeV^{-1}$ lattice spacing.
Choosing the temporal time step as $\delta \tau=0.005 a$ we find that
the violation $\sqrt{|(u^\tau)^2-(u^r)^2-1|}$, summed over all lattice
sites and divided by the number of sites always stays smaller than
one percent, providing yet another check on the numerics.
Finally, we have checked that choosing $a=1$ or $0.5$ $\rm GeV^{-1}$ does
not result in any noticeable deviations of our calculated temperature 
profile, thus we have some confidence that our results are not
strongly affected by numerical artifacts.

\begin{figure}
\vspace{0.6cm}
\begin{center}
\includegraphics[width=0.6\linewidth]{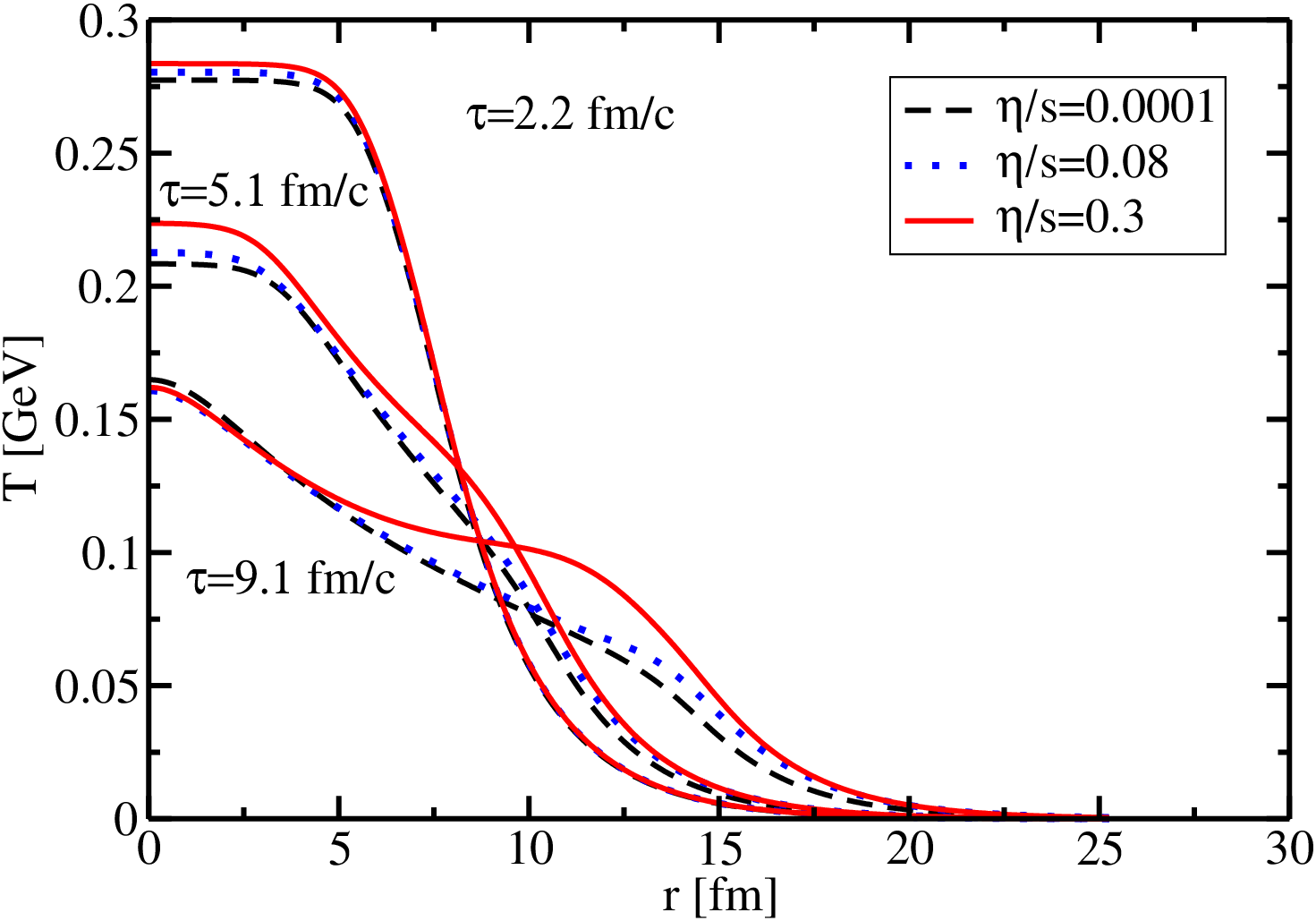}
%\hspace*{1cm}
%\includegraphics[width=0.4\linewidth]{sonic2_2.eps}
\end{center}
\caption{Temperature profile for calculations with different $\eta/s$
(dashed, dotted and solid lines, respectively) for  
three different times (see text for details). 
As expected, for larger values of $\eta/s$, differences to ideal
hydrodynamics are biggest and viscous hydrodynamics
initially cools slower than ideal hydrodynamics. 
However, note that in certain regions and at later times,
viscous hydrodynamics turns out to give temperatures slightly smaller 
than the corresponding ideal hydrodynamic calculation.}
\label{fig:tprof} 
\end{figure}

For early times, the behavior shown in Fig.\ref{fig:tprof} shows that
-- as in the case of neglecting transverse flow \cite{Muronga,Baier:2006um} -- 
in viscous hydrodynamics the temperature decreases slower 
than in ideal hydrodynamics. However, at late times this behavior 
is seemingly counteracted by viscous radial dynamics: at very small 
values of the radius, the viscous hydrodynamic equations result in 
slightly lower
temperatures than in the ideal hydrodynamic case. Note that
such behavior has been also found in 
\cite{Teaney:2004qa,Chaudhuri:2006jd}.

\subsection{Particle Spectra in Viscous Hydrodynamics}

The success of the hydrodynamic picture in
the context of heavy-ion collision builds upon the ability to
fit the particle spectra observed in these collisions.
While now methods of how to convert
hydrodynamic quantities such as energy density and fluid velocity
into particle spectra have reached some sophistication, 
the main building block still seems to
involve the Cooper-Frye freeze-out prescription \cite{CooperFrye}
in some form or the other,
which states that the spectrum of particles
with energy $E$ and momentum $p$ is given by
\beq
E \frac{d^3 N}{d^3p}=\frac{d}{(2\pi)^3} \int p_\mu d\Sigma^\mu
f\left(\frac{p_\mu u^\mu}{T}\right),
\label{specdef}
\eeq
where $d$ is the degeneracy of the particles and $u^\mu$ is the velocity
which comes out of the solution of the hydrodynamic equations.
Here $f$ is the distribution function which -- including 
viscous corrections -- can be written as \cite{Teaney, Baier:2006um} 
\beq
f=f_0\left(1+\frac{p_\mu p_\nu \Pi^{\mu \nu}}{2 T^2 (\epsilon+p)}\right),
\eeq
where for simplicity we take the equilibrium distribution $f_0$ to be
given by the Boltzmann-distribution\footnote{
It is easy to change this to Bose-Einstein or Fermi-Dirac
distributions, but for massive particles we have found the differences
in resulting observables to be minimal.}
$f_0(x)=\exp(-x)$.

Furthermore, $d\Sigma^\mu$ is the normal vector on the freeze-out surface,
parametrized as $ 
d\Sigma^\mu=\left(\cosh{\eta},\cos{\phi}\frac{d
\tau_f(r)}{d\, r}, \sin{\phi}\frac{d \tau_f(r)}{d\, r},
\sinh{\eta}\right) r dr \tau_f(r) d\phi d\eta$ in our choice of
coordinates \cite{Muronga:2004sf}. 
Here $\tau_f(r)$ is the freeze-out time parametrized as a function of
$r$ or -- put differently -- the time at which the slab of matter at
radius $r$ has reached the freeze-out condition.

For this exploratory study, we will apply the Cooper-Frye
freeze-out prescription to convert the hydrodynamic variables
at a single specific temperature (the freeze-out
temperature $T_f$) into transverse momentum spectra for particles.
This is what also has been used
in early ideal hydrodynamic calculations \cite{VonGersdorff:1986yf,McLerran:1986nc}.
Since we use a gluonic equation of state and 
do not include a realistic matching to hadronic degrees of freedom,
we contend ourselves to study the effects of viscosity on
the spectrum of gluons mostly.

%For this exploratory study we will apply the Cooper-Frye
%freeze-out prescription at a single specific temperature (the freeze-out
%temperature $T_f$),
%assuming particles not to interact after this point.
%This is what also has been used
%in early ideal hydrodynamic calculations \cite{VonGersdorff:1986yf,McLerran:1986nc}
%and -- although having been replaced by more complicated freeze-out
%mechanisms recently -- provides us with a rough estimate of the resulting
%particle spectra.
%Furthermore, for simplicity 
%we do not include effects from the hadronic decays; while
%this will prohibit us from matching e.g. the pion-spectra, the slope
%(but not the normalization) 
%of more massive particles (e.g. kaons and protons) does not seem
%to be much influenced by this feed-down mechanism
%\cite{Schnedermann:1993ws}. Thus, by matching
%to the slope of these heavier particles, we hope to be able to circumvent
%somewhat the problems of neglecting the decays.
%Finally, lacking the ability of matching the normalization of the
%hadron spectra we reduce the freedom in the choice of initial conditions by 
%starting all hydrodynamic calculations at a fixed $\tau_0=1$ fm/c.

For the spectrum, in terms of particle transverse 
momentum $p_\perp$, angle $\phi_p$ 
and rapidity $y$ one thus finds
\beq
p_\mu d\Sigma^\mu=\left(m_\perp \cosh(\eta-y)-p_\perp
\cos(\phi-\phi_p) \frac{d \tau_f(r)}{d\, r}\right)
\tau_f(r) r dr d\phi d\eta,
\eeq
with $m_\perp=\sqrt{p_\perp^2+m_0^2}$ and $m_0$ the rest mass of the particle.

\begin{figure}
\vspace{0.6cm}
\begin{center}
\includegraphics[width=0.4\linewidth]{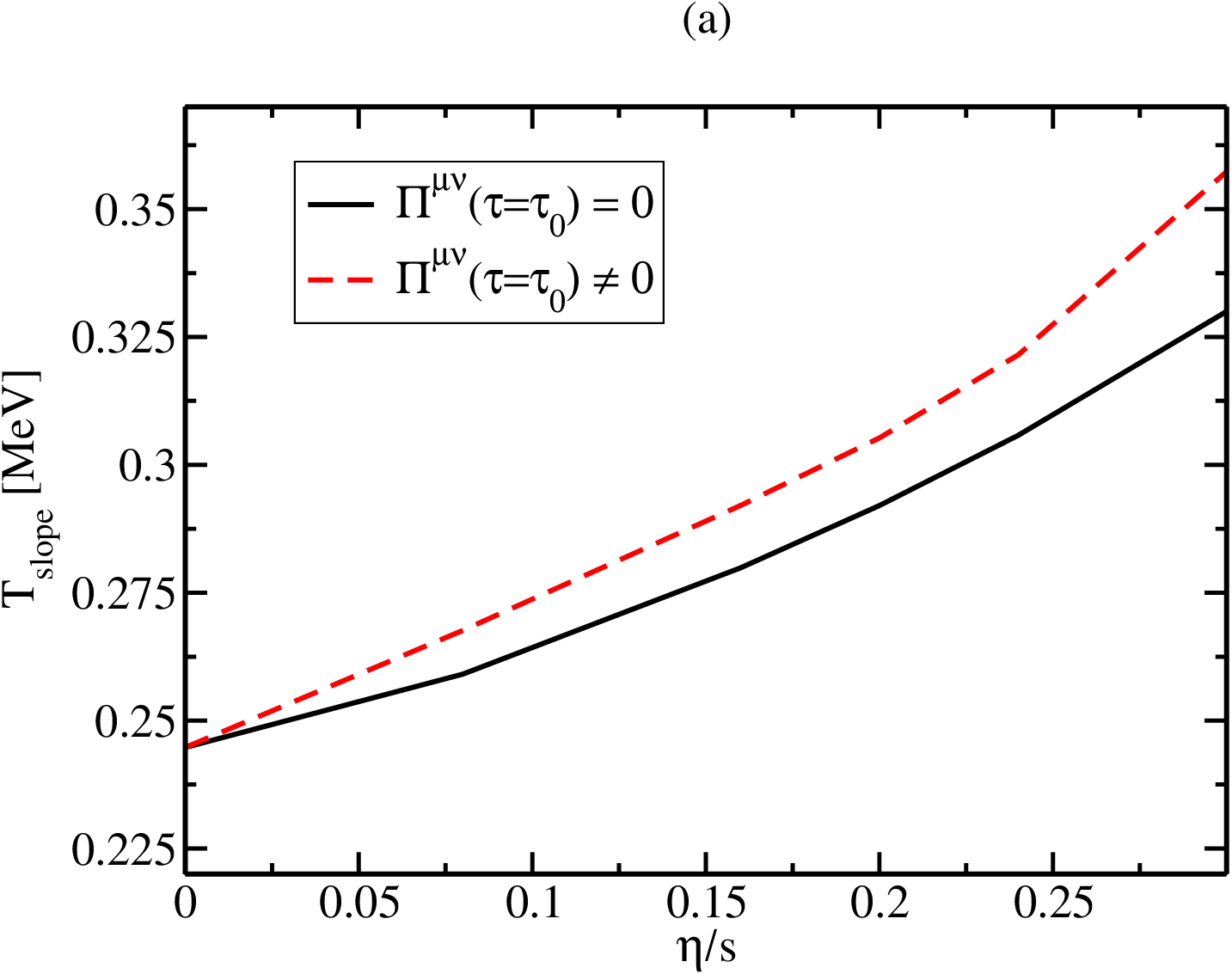}
\hspace*{1cm}
\includegraphics[width=0.4\linewidth]{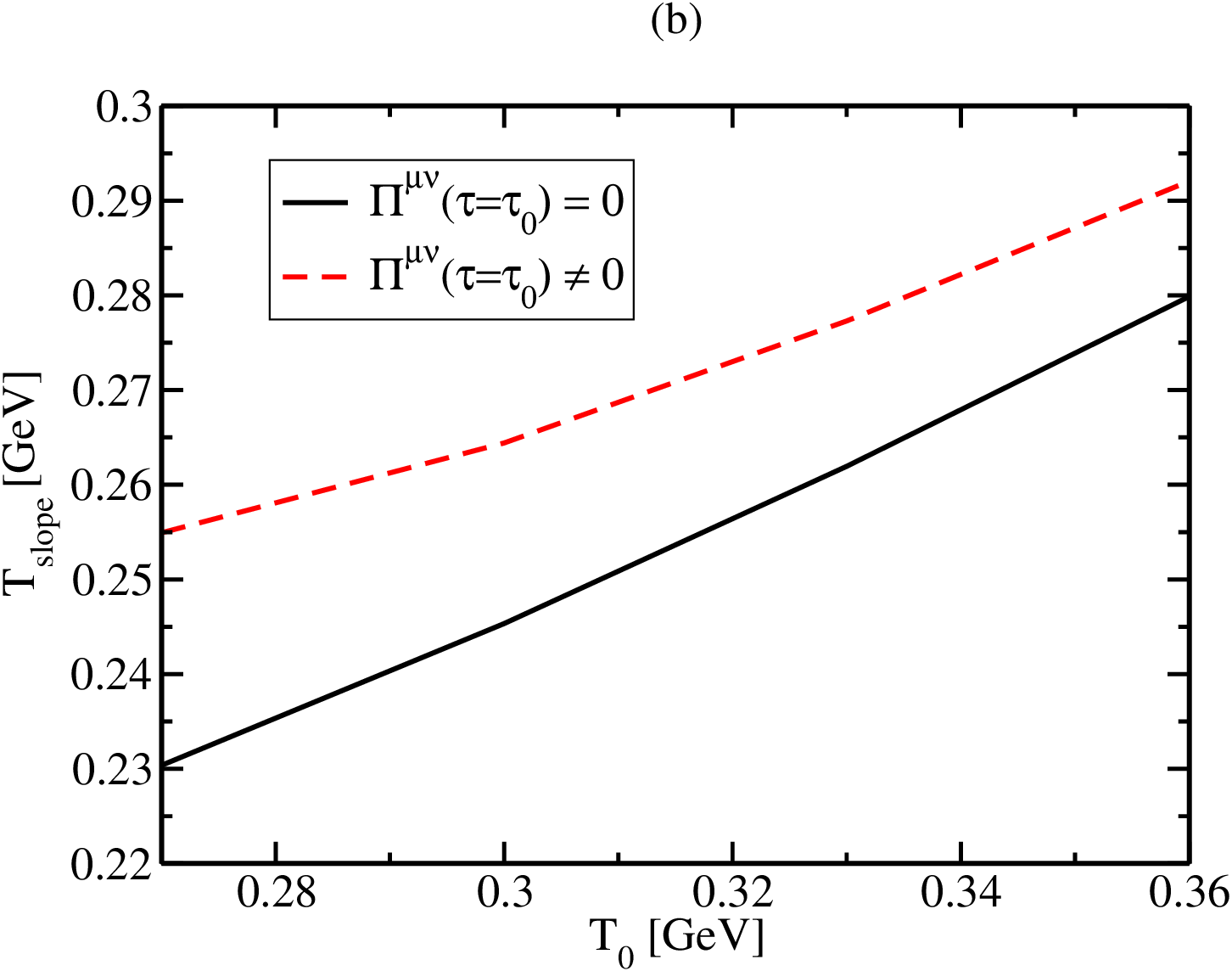}
\end{center}
\caption{Inverse slope parameter $T_{\rm slope}$ for $T_f=0.135$ GeV and 
$\tau_0=1$ fm/c. Two initial conditions for $\Pi^{\mu\nu}$, corresponding
to pressure isotropy (full line) and vanishing longitudinal pressure
(dashed line) at $\tau=\tau_0$ are shown. Choosing $T_0=0.36$ GeV (left),
the spectra become increasingly flatter when raising $\eta/s$, while
this effect can be compensated by lowering $T_0$ (right, shown 
for $\eta/s=0.16$).}
\label{fig:gluspec}
\end{figure}

\begin{figure}
\vspace{0.6cm}
\begin{center}
\includegraphics[width=0.6\linewidth]{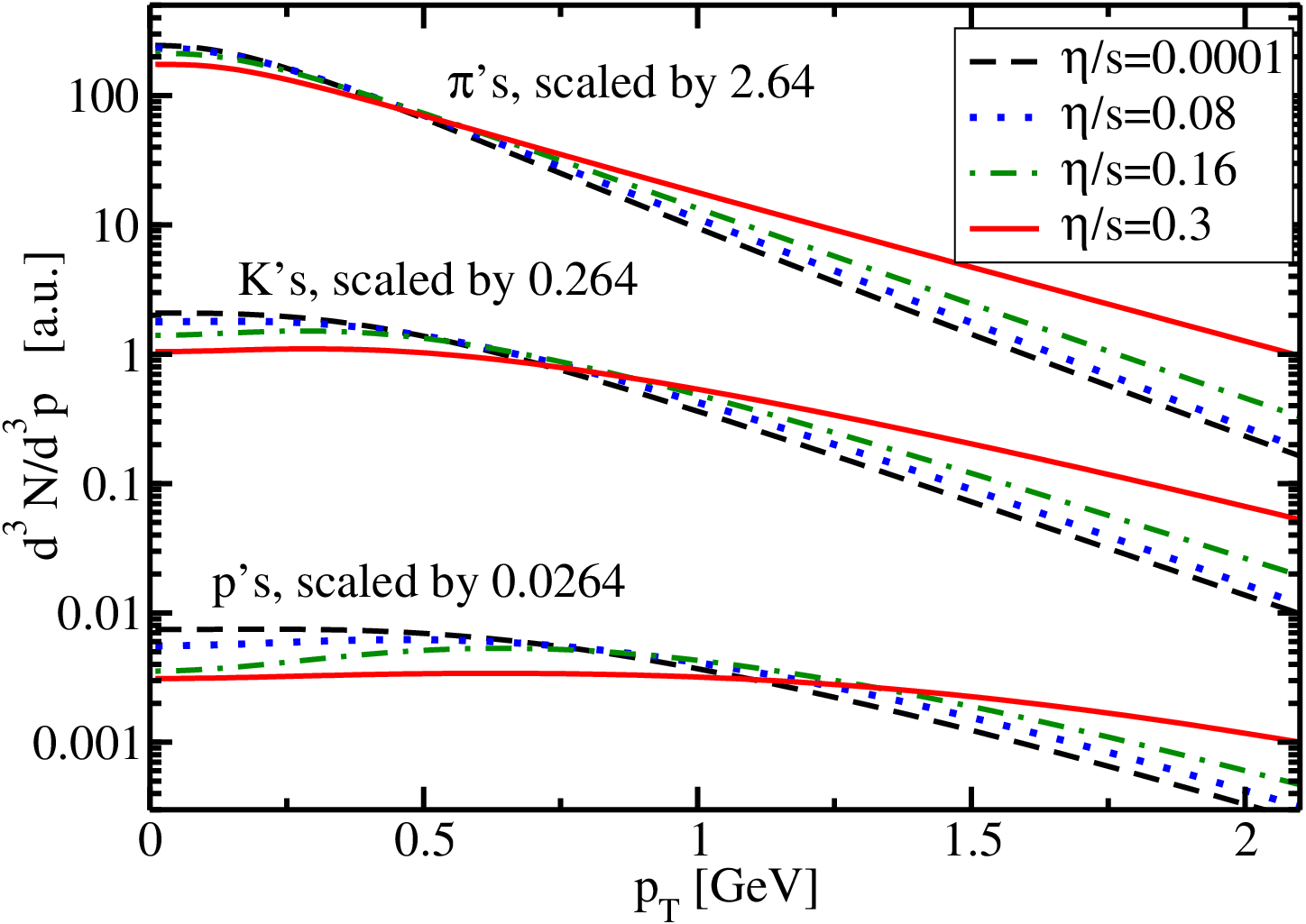}
\hspace*{1cm}
\end{center}
\caption{Mass dependence of viscous effects: shown are
spectra for pions, kaons and protons for a freeze-out
temperature of $T_f=0.135$ GeV and for $T_0=0.36$ GeV and $\Pi^{\mu\nu}=0$
at $\tau_0=1$ fm/c. As can be seen, the higher the mass of the particle,
the stronger do viscous effects change its low $p_\perp$ behavior.}
\label{fig:spectra} 
\end{figure}

Since in our calculations we are only including transverse flow we have
\beq
p_\mu u^\mu=\left(m_\perp \cosh(\eta-y) u^\tau-p_\perp \cos(\phi-\phi_p) u^r\right),
\eeq
which allows us to integrate out both angles $\phi$ and rapidities
$\eta$ in Eq.(\ref{specdef}).
Using
\bqa
\int_{-\infty}^{\infty}d\eta \cosh^n{\eta} \exp(- x \cosh{\eta}) &=&
\left(-\frac{d}{d\, x}\right)^n 2 K_0(x),\nonumber\\
\int_0^\pi \frac{d\phi}{\pi} \cos{n \phi} \exp{(x \cos{\phi})}&=&I_n(x),
\eqa
where $K_n(x)$ and $I_n(x)$ are modified Bessel functions
one finds for the particle spectrum
\beq
E \frac{d^3 N}{d^3p}=E \frac{d^3 N_0}{d^3p}+E \frac{d^3\, \delta N}{d^3p},
\label{splitting}
\eeq
with the equilibrium part taking the form
\beq
E \frac{d^3 N_0}{d^3p}=\frac{2 d}{(2\pi)^2} \int r dr \tau_f(r)
\left[m_\perp I_0(u^r p_\perp/T) K_1(u^\tau m_\perp/T)-
\frac{d\tau_f(r)}{d\,r} p_\perp I_1(u^r p_\perp/T) K_0(u^\tau m_\perp/T
)\right],
\label{Neq}
\eeq
where the integral over $r$ runs from $0$ to the maximum freeze-out
radius if $\tau_f(r)$ is a single-valued function (else one has to introduce a
different parametrization of the freeze-out surface).
Noting that the Bessel $K$ functions always have the argument
$u^\tau m_\perp/T$ (and similarly for the $I$'s) we refrain from writing the
argument in the following.
Noting that 
\bqa
p_\mu p_\nu \Pi^{\mu \nu}&=& \Pi^r_r \left[2 v m_\perp p_\perp
\cosh(y-\eta) \cos(\phi_p-\phi)-p_\perp^2 \cos(2
(\phi_p-\phi))\right.\nonumber\\
&&\left.-v^2 m_\perp^2 \cosh^2(y-\eta)-v^2 p_\perp^2 \sin^2
(\phi_p-\phi) \right]\nonumber\\
&&+\Pi^\eta_\eta \left[-m_\perp^2 \sinh^2(y-\eta)+p_\perp^2 
\sin^2(\phi_p-\phi)\right],
\eqa
we then find the for the dissipative corrections to the spectrum
\bqa
E \frac{d^3\, \delta N}{d^3p} &=&\frac{d}{(2\pi)^2}\int r dr 
\frac{\tau_f(r)}{2 T^2 (\epsilon+p)}
\left\{
2 p_\perp m_\perp v \Pi^r_r \left[m_\perp(K_0+K_2) I_1-p_\perp
\frac{d\tau_f}{d\, r} K_1 (I_2+I_0)\right]
\right.\nonumber\\
&&\left.
-p_\perp^2 \Pi^r_r \left[2 m_\perp K_1 I_2-p_\perp 
\frac{d\tau_f}{d\,r}
K_0 (I_3+I_1)\right]\right.\nonumber\\
&&\left.
-v^2 m_\perp^2 \Pi^r_r \left[\frac{1}{2}m_\perp(3 K_1+K_3) I_0-
p_\perp \frac{d\tau_f}{d\,r} (K_0+K_2) I_1\right]\right.\nonumber\\
&&\left.
-v^2 p_\perp^2 \Pi^r_r \left[m_\perp K_1 (I_0-I_2)-p_\perp 
\frac{d\tau_f}{d\,r} K_0 \frac{1}{2} (I_1-I_3)\right]\right.\nonumber\\
%&&\left.
%+m_\perp^2 \Pi^\eta_\eta \left[m_\perp 2 K_1 I_0-p_\perp 
%\frac{d\tau_f}{d\,r} 2 K_0 I_1\right]\right.\nonumber\\
&&\left.
+p_\perp^2 \Pi^\eta_\eta \left[m_\perp K_1 (I_0-I_2)-p_\perp
\frac{d\tau_f}{d\,r} K_0 \frac{1}{2} (I_1-I_3)\right]\right.\nonumber\\
&&\left.
-m_\perp^2 \Pi^\eta_\eta \left[m_\perp \frac{1}{2} (K_3-K_1) I_0
-p_\perp \frac{d\tau_f}{d\,r} (K_2-K_0) I_1\right]
\right\},
\label{eq37}
\eqa
where we remind that $v=u^r/u^\tau$.

In Fig.\ref{fig:gluspec}, we show the inverse slope parameter $T_{\rm slope}$ 
of gluons which we define by calculating the gluonic
spectrum at $T_f=0.135$ GeV and fitting it by
$$
E\frac{d^3 N}{d^3 p}\sim \frac{1}{T_{\rm slope}^2} 
\exp{[-p_\perp/T_{\rm slope}]} 
$$
for $0.2<p_\perp<1$ GeV (c.f.\cite{Adler:2003cb}). 
In Fig.\ref{fig:gluspec}(a), the slope
has been calculated for $T_0=0.36$ GeV, $\tau=1$ fm/c and for
the two extreme cases
$\Pi^{\mu\nu}(\tau_0)=0$ (full line) and zero initial longitudinal
pressure Eq.(\ref{CGC}) (dashed line).
As can be seen from this figure, increasing $\eta/s$ and leaving
all other parameters unchanged leads to an increasing $T_{\rm slope}$
(``flatter spectra'') 
for gluons, with no dramatic difference between the two different
initial choices for $\Pi^{\mu \nu}$.
However, as
has been anticipated from our earlier studies neglecting
the effect of transverse flow \cite{Baier:2006um},
one can compensate this effect by changing the effective
initial conditions. This can be seen in Fig.\ref{fig:gluspec}(b), 
where we show the spectral slope for the same freeze-out temperature, but
different initial temperatures $T_0$.

It is also interesting to study how the presence of viscosity
affects massive particles. To this end, hypothetical
spectra of pions, kaons and protons for $T_f=0.135$ GeV, 
$T_0=0.36$ GeV and $\Pi^{\mu\nu}(\tau_0)=0$ at $\tau_0=1$ fm/c
are shown in Fig.\ref{fig:spectra}. These spectra cannot be directly
interpreted as real particle spectra because a realistic matching
to a hadronic equation of state and the effects from
higher mass resonance decays  \cite{Schnedermann:1993ws}
are missing in this study\footnote{
See however \cite{Romatschke:2007jx} for a comparison to experimental data.}.
Nevertheless, from Fig.\ref{fig:spectra} one can glean that
the more massive a particle is, the more
viscosity affects its spectrum, in particular at low
$p_\perp$. Indeed, this can be traced back to Eqs.(\ref{Neq},\ref{eq37})
which in the limit of vanishing $p_\perp$ and neglecting radial dynamics
($v=0$) predict negative $ d^3 N/d^3p$ for large $m_0/T$,
more specifically for 
\beq
\frac{m_0}{T}>\frac{2 (\epsilon+p) - \frac{15}{8} \Pi^\eta_\eta}{
\Pi^\eta_\eta}.
\eeq
Thus it seems that -- whenever $\Pi^\eta_\eta/(\epsilon+p)$ becomes
non-negligible -- viscous corrections $\delta N$ to the spectrum
of high-mass particles become very large, e.g. more than 100 percent
at low $p_\perp$. While it is unclear at which value of $\eta/s$
this starts to be a problem in practice, it nevertheless serves
as an indication that the assumption of small deviations
from equilibrium  \cite{Baier:2006um} is breaking down. 
Consequently, the reliability of the tool we have used 
to probe the system dynamics,
namely viscous hydrodynamics, becomes questionable. 
Thus, for $\eta/s$ larger than a critical value, one probably has to use
a different model than hydrodynamics to correctly calculate observables
that are to be compared to experiments.

\section{Conclusions}
\label{sec:five}

We have studied the effect of shear viscosity in a hydrodynamic
description of central heavy-ion collisions. We presented a simple
algorithm to solve the relevant equations numerically and have successfully
carried out several tests on this algorithm. These tests are not specific
to our algorithm, but can in general be used to test any algorithm
for solving relativistic viscous hydrodynamics.
Assuming an ideal equation of state $\epsilon=3 p$ for simplicity we
calculated the time evolution of the
temperature profile of a central heavy-ion collision, finding
that while viscous hydrodynamics in general cools slower, certain regions
at later times
may cool faster than in a corresponding ideal hydrodynamic calculation.

We also calculated the effect of viscosity on the slope of gluon spectra,
finding that for small values of $\eta/s$, changes can largely be compensated
by lowering the temperature at which the hydrodynamic evolution is started.
For massive particles we find that viscosity changes the spectrum
the more the higher the mass of the particle under consideration.
We give arguments that for a sufficiently large value of $\eta/s$,
%this indicates that 
corrections which 
in the derivation of the viscous hydrodynamic equations had been assumed to
be small actually become large, thus signaling the possible breakdown of
any hydrodynamic description of the system.
%such that the slopes are in fair agreement with experimental data.
%At values of $\eta/s=0.3$ or larger, however, experimental data seems not to be
%well described by viscous hydrodynamics anymore. We show that
%at these values of $\eta/s$, viscous corrections to the spectra which
%in the derivation of the viscous hydrodynamic equations had been assumed to
%be small actually become large, thus signaling the possible breakdown of
%any hydrodynamic description of the system.

Even though our simplifying assumptions (ideal equation of state, no
feed-down correction, only radial flow) leave ample room for improvement,
we hope that our study provides the basis for coming viscous
hydrodynamic algorithms as well as fits to experimental data.

\acknowledgments

We would like to thank U.W.~Heinz, P.~Huovinen, J.Y.~Ollitrault and 
D.~Rischke for 
illuminating discussions on ideal hydrodynamics.
Also, we want to thank an anonymous referee for several constructive remarks.
PR was supported partially by BMBF 06BI102 and the US Department of Energy,
grant number DE-FG02-00ER41132.

\end{document}